\documentclass[prd,twocolumn,amsmath,amssymb,showpacs,floatfix,nofootinbib,preprintnumbers,superscriptaddress]{revtex4-2}
\DeclareUnicodeCharacter{2009}{\,}
\usepackage{graphics,epsfig,psfig}
\usepackage{todonotes}
\usepackage[normalem]{ulem}
\usepackage{xcolor}
\usepackage{float}
\usepackage{dcolumn}
\usepackage{kantlipsum}
\usepackage[style=base, singlelinecheck=off, font=small, labelfont=bf, 
justification=raggedright]{caption}
\usepackage{subfigure}
\usepackage[]{inputenc,amssymb}
\usepackage{natbib}
\usepackage{url}
\usepackage{orcidlink}
\usepackage{amsmath}
\graphicspath{{plots_arxiv/}}
\usepackage{cancel}

  \makeatletter
\def\@fnsymbol#1{\ensuremath{\ifcase#1\or \dagger\or \ddagger\or
   \mathsection\or \mathparagraph\or \|\or **\or \dagger\dagger
   \or \ddagger\ddagger \else\@ctrerr\fi}}
    \makeatother
\def \be{\begin{equation}}
\def \ee{\end{equation}}
\def \bea{\begin{eqnarray}}
\def \eea{\end{eqnarray}}

\definecolor{webgreen}{rgb}{0,.5,0}
\definecolor{webbrown}{rgb}{.6,0,0}

\newcommand{\icarogw}{\texttt{icarogw}}
\newcommand{\bilby}{\texttt{Bilby}}

\begin{document}


\title{Blinded Mock Data Challenge: Is the Spectral Siren Technique Robust for Measuring the Hubble Constant?}

\author{Christos Karathanasis\,\orcidlink{0000-0002-0642-5507}}\altaffiliation{Simulation team member}\affiliation{Institut de Física d’Altes Energies (IFAE), Barcelona Institute of Science and Technology, Barcelona, Spain}
\author{Suvodip Mukherjee\,\orcidlink{0000-0002-3373-5236}}\altaffiliation{Editorial team, simulation and coordinator}\affiliation{Department of Astronomy and Astrophysics, Tata Institute of Fundamental Research, Mumbai 400005, India}\email{suvodip@tifr.res.in}
\author{Lalit Pathak\,\orcidlink{https://orcid.org/0000-0002-9523-7945}}\altaffiliation{Editorial team and simulation team}\affiliation{Department of Astronomy and Astrophysics, Tata Institute of Fundamental Research, Mumbai 400005, India}\affiliation{Inter-University Centre for Astronomy and Astrophysics (IUCAA), Pune 411007, India}
\author{Sergio Vallejo-Pe\~na\, \orcidlink{0000-0002-6827-9509}}\altaffiliation{Editorial team and Analysis team}\affiliation{Instituto de Fisica, Universidad de Antioquia, A.A.1226, Medellin, Colombia}
\author{Mohit Raj Sah\,\orcidlink{0009-0005-9881-1788}}\altaffiliation{Editorial team and simulation team}
\affiliation{Department of Astronomy and Astrophysics, Tata Institute of Fundamental Research, Mumbai 400005, India}
\author{Beno\^it Revenu\,\orcidlink{0000-0002-7629-4805}}\altaffiliation{Simulation team}\affiliation{Subatech, CNRS - Institut Mines-Telecom Atlantique - Universit\'e de Nantes, France}\affiliation{Université Paris-Saclay, CNRS/IN2P3, IJCLab, 91405 Orsay, France}
\author{Antonio Enea Romano\,\orcidlink{0000-0002-0314-8698}}\altaffiliation{Editorial Team and Analysis team}\affiliation{Instituto de Fisica, Universidad de Antioquia, A.A.1226, Medellin, Colombia}
\author{Juan Garcia-Bellido\,\orcidlink{0000-0002-9370-8360}}\affiliation{Instituto de Fisica Teorica, Universidad Autonoma de Madrid, Cantoblanco 28049 Madrid}
\date{\today}

\begin{abstract}
The measurement of the Hubble constant from gravitational wave (GW) sources is one of the independent avenues to shed light on the Hubble tension, which is associated with about an $8\%$ mismatch in the value of the Hubble constant inferred from low-redshift and high-redshift cosmological probes. Such a key measurement is expected from GW sources as it is a direct measurement of the Hubble constant using the luminosity distance without the need for any luminosity distance calibration. However, such a measurement relies strongly on the reliability of the independent inference of the source redshift of the GW source. As a result, it becomes pertinent to gauge the accuracy and precision of techniques in understanding their reliability in inferring redshifts of GW sources. In this work, we show the requirement of the spectral siren technique in knowing the mass distribution of BBHs across cosmic redshifts in order to make a reliable inference of the Hubble constant. We show by a blinded mock data challenge analysis the criticality in capturing the underlying metallicity dependence of the BBH mass distribution and its interplay with time-delay distribution for a robust inference of the Hubble constant using the spectral siren technique. In order to have a reliable measurement of the Hubble constant at the level required to resolve the Hubble tension in the future, the mass distribution of the BBHs needs to be independently inferred at all relevant redshifts with an accuracy less than the statistical uncertainty. Otherwise, a mismatch of the true model and the underlying assumption made in the analysis can lead to a best-fit model for the wrong value of both BBH population parameters as well as the Hubble constant. 
\end{abstract}

\maketitle

\section{Introduction}\label{sec:Intro}

The detection of gravitational waves (GWs) from binary compact object mergers, such as binary neutron stars (BNSs) and binary black holes (BBHs), has opened a new observational window for the Universe, enabling us to study Cosmology. One of the key cosmological studies possible from the GW observations is in the inference of the expansion history of the Universe, which can shed light on the energy content of the Universe and the local expansion rate, the Hubble constant of the Universe\cite{1929PNAS...15..168H}. Though there are several precise observational techniques to measure the Hubble constant using electromagnetic observations, such as cosmic microwave background (CMB), baryon acoustic oscillation (BAO), supernovae, lensing, etc., gravitational wave (GW) sources bring a unique advantage to measure the luminosity distance without any external calibration \cite{Schutz}. As a result, they are also called the standard sirens. 

Such a standard siren measurement can provide accurate and direct inference of the Hubble constant, if the redshifts can be inferred directly. The inference of the redshifts for the GW sources can be possible in multiple ways, such as directly by identifying the host galaxy and its redshift for GW sources with electromagnetic counterparts, such as bright standard siren GW170817 \cite{Abbott:2017xzu}, or also by statistical techniques for GW sources without electromagnetic counterparts, such as cross-correlation technique\cite{Mukherjee:2018ebj,Mukherjee:2019wcg,Mukherjee:2020hyn, Bera:2020jhx,Mukherjee:2022afz,Ferri:2024amc, Afroz:2024joi}, host galaxy identification technique\cite{Schutz, Fishbach:2018gjp,Soares-Santos:2019irc,PhysRevD.101.122001, Soares-Santos:2019irc, Palmese:2021mjm, LIGOScientific:2021aug, Gray:2023wgj, Hanselman:2024hqy,LIGOScientific:2025jau}, and spectral-siren technique\cite{Farr:2019twy, Mastrogiovanni:2021wsd, Mukherjee:2021rtw, Leyde:2022orh, LIGOScientific:2021aug, Ezquiaga:2022zkx, Mastrogiovanni:2023zbw, Leyde:2023iof, Pierra:2023deu, MaganaHernandez:2024uty, Farah:2024xub, Mali:2024wpq,LIGOScientific:2025jau}. Though each of these techniques can provide a Hubble constant measurement, the accuracy of such a measurement depends on the robustness of the technique in inferring the Hubble constant. As the key advantage of GW cosmology is that it can provide an independent and accurate measurement of the Hubble constant, it is pertinent to understand the robustness of the techniques used. In light of this, a previous blinded mock data challenge (\texttt{Blinded-MDC}) investigating a phenomenological redshift-dependent mass model was performed \cite{agarwal2025blinded}, where the metallicity dependence and the delay time distribution were not included. Several other analyses were also carried out to understand the robustness of the spectral-siren technique \cite{Mukherjee:2021rtw, Pierra:2023deu}.  

In this work, we focus on the critical assessment of the spectral-siren technique using an astrophysically motivated scenario of a population with two critical aspects: (i) inclusion of the delay-time distribution in the BBH population model, (ii) inclusion of stellar metallicity dependence on the PISN mass-gap. Both these effects together lead to a metallicity-dependent BBH merger rate per unit cosmic volume per unit stellar mass. As there is a redshift evolution of the metallicity distribution in the Universe, which depends on several astrophysical phenomena, it becomes important to understand the impact of mis-modelling of astrophysical BBH population on the spectral siren technique.   

{It is important to note that the issue of mis-modelling of astrophysical BBH population for inferring cosmological results can be broadly classified into two aspects: (i) ignorance of the true BBH population model across cosmic time and (ii) ignorance of the accurately inferred value of a particular model. For the first scenario, the misunderstanding lies in the model assumption, such as the power-law model, broken power-law model, power-law+Gaussian model, power-law+multi-peak model, or maybe some other BBH population model. This will be clearer with more data in the future. The true class of the BBH population model for different cosmic times is still unknown. The second scenario is related to accurately capturing these models by either parametric or non-parametric inference techniques. Mismatch in the inference of these values, which can characterize the models, can also cause errors in the inference of the Hubble constant. In this \texttt{Blinded-MDC} analysis, we are focusing on the redshift evolution of only a particular class of model, the power-law+Gaussian (PLG) model, and the redshift evolution of this specific class of model. The goal of this \texttt{Blinded-MDC} is to check for a particular class of injected model (which we have considered PLG in this analysis), how robust is the inference of the cosmic expansion history when there is redshift evolution of the BBH population in the injection, but it is not captured in the analysis. However, in reality, the underlying assumed PLG model can be different and can cause an additional layer of uncertainty. So, in order to have a robust measurement of cosmological parameters, one would need a robust measurement of the BBH population model across cosmic time, which is accurately inferred from observations.} 


In order to perform a quantitative test of the robustness of this technique, we perform the \texttt{Blinded-MDC} of the spectral-siren technique for a setup where the underlying astrophysical model of BBH is unknown for the flat Lambda Cold Dark Matter (LCDM) model for O4 sensitivity of the LIGO-Virgo-KAGRA Collaboration \cite{LIGOScientific:2014pky, VIRGO:2014yos, PhysRevD.88.043007}.  {Our results point out, for the first time, using a \texttt{Blinded-MDC} that the impact of the metallicity-dependent BBH formation and evolution scenario on the Hubble constant inference using the spectral-siren technique, and how the degeneracy between parameters can drive the inferred value of the Hubble constant away from the true one.}   As a result, in order to resolve the Hubble tension \cite{Verde:2019ivm, DiValentino:2021izs}, which is about $8\%$ difference in the value of the Hubble constant inferred using CMB at around $H_0=67$ km/s/Mpc \cite{Planck:2018vyg} and from low redshift probes at around $H_0=73$ km/s/Mpc \cite{Riess:2019cxk}, one would need to independently know the BBH mass distribution with an accuracy much less than $8\%$.  {So, in the regime when the statistical uncertainty on the Hubble constant will be subdominant over the systematic errors, it would be crucial to understand these sources of systematics to resolve the Hubble tension.} 

The findings of the paper are organised as follows. In Sec. \ref{sec:setup} and Sec. \ref{sec:case_study_bmdc}, we discuss the setup of the MDC and the simulation suite considered in the analysis, Sec. \ref{sec:methods} discusses the Bayesian framework used for the inference, and Sec. \ref{sec:results} and Sec. \ref{conc} discuss the findings from the \texttt{Blinded-MDC} and its conclusion on standard siren cosmology.

\section{Setup for the Blinded Mock Data Challenge}
\label{sec:setup}

The aim of the \texttt{Blinded-MDC} presented in this work is to quantify the systematic bias in the inference of the cosmological parameters that arises when the underlying population of binary BHs differs from the simplified assumptions made in the cosmological analysis. This work represents the second analysis in our series of blinded mock data challenges. In the first study, we examined the impact of a redshift–dependent BH mass distribution on the inference of cosmological parameters \citep{agarwal2025blinded}. In the present analysis, we extend this framework by allowing the merger rate to follow a time–delay–modulated Madau–Dickinson-like star–formation rate \citep{Madau:2014bja} in addition to the redshift–dependent mass distribution due to metallicity evolution. This combination of a redshift-evolving mass spectrum and a delayed merger-rate model provides a more complete description of the underlying BH population and allows us to assess the resulting systematics in a controlled, blinded setting. {The fiducial BBH population model considered in this analysis is based on the PLG model with redshift dependence for the inference and the PLG model for analysis. As a result, we are considering the underlying type of model, still the same PLG for both simulation and analysis, and only allowing redshift evolution of the parameters for the PLG model. However, it is important to note that the actual BBH population in nature is still not known accurately. Recent observations using GWTC-4 have shown support for multiple peaks in the BBH population model \cite{LIGOScientific:2025pvj}. So, it would be interesting to explore how the difference in the underlying simulation model and analysis model would impact the Hubble constant inference. As we still do not know the true underlying BBH population model, this \texttt{Blinded-MDC} analysis, which considers a similar type of model (PLG model) for both injection and simulation, is still following simplistic assumptions. In reality, there can be modeling errors that can propagate to uncertainty in the Hubble constant inference. In the future, these kinds of studies would be essential to gauge the robustness of the Hubble constant measurement using GW sources.}

The MDC is performed in a blinded manner to prevent any kind of confirmation bias or bias in the analysis setup. The blinding analysis follows the same procedure as in previous MDC \citep{agarwal2025blinded}: a \textbf{simulation team} generates the gravitational-wave events using the chosen astrophysical and cosmological parameters, while the \textbf{analysis team} receives only the parameter estimation posteriors for each detected event. No information about the injected population hyperparameters is disclosed to the analysis team; they are informed only of the general functional form of the population model until all analyses are completed. The analysis team tries to recover the hyperparameters using the fiducial assumption of redshift and metallicity-independent BBH mass distribution with no time delay.

\textit{\textbf{Blinded-MDC Simulation setup}}: The mock data are generated by the simulation team using the \texttt{GWSIM} package\cite{karathanasis2023gwsim}, which incorporates the appropriate duty cycles and noise power spectral densities. For each realization, the astrophysical and cosmological hyperparameters are drawn randomly from broad prior ranges to ensure full coverage of the allowed population space. These sampled parameters are then supplied to \texttt{GWSIM} \citep{karathanasis2023gwsim}, which simulates a complete catalog of BBH mergers. Events are retained only if their network matched–filter signal–to–noise ratio, computed across the LIGO-Hanford, LIGO-Livingston, and Virgo detectors, exceeds the threshold $\rho_{\mathrm{th}}$. In this analysis, all injections are generated using an SNR threshold of $\rho_{\mathrm{th}} = 12$ with the O4 noise power spectral density (PSD) \citep{abbott2016observation,Abbott_2021,Abbott_2019,LIGOScientific:2021djp,abac2025gwtc}, and adopting a detector duty cycle of 0.75, independent between detectors.  The injections are then passed through parameter estimation using the package \texttt{Bilby} \citep{bilby_paper}, yielding posterior samples for the detector-frame source parameters. The analysis team receives only these posterior samples. All injected population and cosmological parameters remain hidden until the completion of the analysis.

\textit{\textbf{Blinded-MDC Analysis setup}}: The analysis is carried out using the \texttt{icarogw} code \citep{Mastrogiovanni:2023zbw}, which enables a joint hierarchical inference of the cosmological and population hyperparameters from the parameter–estimation posteriors of the detected events. The analysis team receives only the posterior samples for each of the detected GW events. The team conducts the hierarchical inference under fixed astrophysical modelling assumptions that intentionally differ from those used to generate the mock events. In particular, the analysis assumes a redshift–independent Power Law + Gaussian mass distribution and a merger–rate parametrization without any explicit time delay.  

Once the analysis is fully completed and all consistency tests are satisfied, the true injected parameters are unblinded. This enables a direct comparison between the inferred and injected quantities, allowing us to quantify systematic biases arising from mismodelling of the underlying astrophysical population. In the remainder of the paper, we describe the MDC setup in detail and present the results of the injection and analysis.

\section{Simulated BBH Population Model}
\label{sec:case_study_bmdc}

In the first MDC paper, the \texttt{Blinded-MDC} framework investigates systematic errors arising from astrophysical mis-modeling due to the redshift dependence of the mass model. In the present work, in addition to the redshift-dependent mass model, the injected model also includes the delay-time distribution in the merger rate. We use the simulation code \texttt{GWSIM} \citep{karathanasis2023gwsim} for generating the mock sources with O4 sensitivity, and the parameter estimation for all simulated GW events is carried out using the \texttt{bilby} framework~\citep{bilby_paper}, employing the \texttt{IMRPhenomPv2} waveform approximant~\citep{waveform1,waveform2,waveform3}. We adopt the standard \texttt{bilby} BBH priors, with all spin components fixed to zero (i.e.\ delta-function spin priors). The posterior is obtained for nine detector-frame parameters: the chirp mass $\mathcal{M}$, the mass ratio $q = m_2/m_1 \leq 1$, the luminosity distance $d_L$, the sky position (right ascension RA and declination Dec), the inclination angle $\iota$, the polarization angle $\Psi$, the phase $\phi$ of the gravitational-wave signal, and the coalescence time $t_c$. Given the detector noise levels relevant for this study, the restriction to a non-spinning waveform model is not expected to introduce any noticeable bias in the cosmological parameters inferred from the mock dataset.

\textbf{\textit{Black Hole population model}}: We assume black holes have an astrophysical origin and consequently model the binary merger rate $R(z)$, using a functional form motivated by the cosmic star formation rate (SFR), adopting the Madau–Dickinson parametrization \cite{Madau:2014bja}.
\begin{equation}
\label{eq:madau}
    R_{\rm SFR}(z) \propto  (1+z)^{\gamma} \frac{1+(1+z_p)^{-(\gamma+\kappa)}}{1+(\frac{1+z}{1+z_p})^{(\gamma+\kappa)}},
\end{equation}
{where $\gamma$ and $\kappa$ control the low and high-redshift slopes, respectively, and $z_p$ denotes the redshift at which the SFR peaks. In this analysis, the injected values of $\gamma$, $\kappa$, and $z_p$ are fixed to the best-fit values from \cite{Madau:2014bja}.}

The formation of a BBH system is not instantaneous: massive stars require several million years to evolve into compact remnants, and the subsequent processes of binary hardening, orbital decay, and eventual merger can span from tens of millions to several billions of years \citep{dominik2012double,mapelli2017cosmic,vitale2019measuring,boesky2024binary}. To account for this astrophysically motivated lag between the birth of the progenitor system and the merger of the resulting BHs, we incorporate a time-delay distribution into the function $R_{\rm SFR}(z)$. This delay-time prescription effectively links the binary formation epoch to the merger redshift, providing a more realistic description of the cosmic evolution of the BBH population \citep{Karathanasis:2022rtr,karathanasis2023gwsim}.

\begin{equation}
    R(z) = R_0\frac{\int_{z}^{\infty} P_{t}(t_d\mid t^{\text{min}}_d, t^{\text{max}}_d, d) R_{\text{SFR}}(z_f)\frac{dt}{dz_f}dz_f}{\int_{0}^{\infty} P_{t}(t_d\mid t^{\text{min}}_d, t^{\text{max}}_d, d) R_{\text{SFR}}(z_f)\frac{dt}{dz_f}dz_f}
\end{equation}

where the delay-time distribution is given by
\begin{equation}
    P_{t}(t_d\mid 
     t^{\text{min}}_d, t^{\text{max}}_d, d) \propto 
    \begin{cases}
        t^{-d}_d & ,\text{for}\,\,\, t^{\text{min}}_d < t_d < t^{\text{max}}_d,\\
        0, &\text{otherwise}
    \end{cases}
\end{equation}

For the mass distribution, we adopt the Power-law-Gaussian (PLG) model for the primary mass of BBH and a simple conditional power-law for the secondary mass, given as \citep{KAGRA:2021duu,karathanasis2023gwsim,abac2025gwtc}

\begin{widetext}
\begin{align}
    P_{s_1}(m_1|z) =
    \begin{cases}
        (1-\lambda(z)) \mathcal{P}(m_1,-\alpha(z)) + \lambda(z) ~ \mathcal{G}(m_1, \mu_g(z),\sigma_g(z)), & m_{\rm min} < m_2 < m_{\rm max},\\
        0, &\text{otherwise}
    \end{cases}
\label{Pz1}
\end{align}

\begin{align}
    P_{s_2}(m_2|m_1,z) =
    \begin{cases}
        (1-\lambda(z)) \mathcal{P}(m_1,-\beta(z)), &m_{\rm min} < m_2 < m_1,\\
        0, &\text{otherwise}.
    \end{cases}
\label{Pz1}
\end{align}
\end{widetext}
\twocolumngrid

We express the redshift dependence of all parameters as $x(z) = x_0 + z \epsilon_{x}$, where $x_0$ represents the parameter value at redshift $z=0$ \citep{karathanasis2023gwsim}. This linear redshift dependence is applied uniformly to all mass distribution parameters. The astrophysical and cosmological hyperparameters used to obtain the events are given in Tab.~\ref{tab:injections}.

The redshift evolution of the mass distribution can arise from the effects of pair-instability supernovae (PISN), which play an important role in shaping the high-mass end of the stellar-origin black hole population \citep{heger2002nucleosynthetic,kasen2011pair,farmer2019mind}. The mass scale affected by PISN depends sensitively on stellar metallicity \citep{farmer2019mind}: low-metallicity stars lose less mass through winds and can therefore produce heavier remnants \citep{mokiem2007empirical,van2005metallicity,vink2001mass}. Since metallicity decreases with redshift \citep{10.1007/978-1-4020-9190-2_9,mannucci2010fundamental}, the characteristic mass scale influenced by PISN physics is expected to increase at higher redshift, making the BH mass distribution inherently redshift dependent.  When combined with a non-zero delay-time distribution between stellar formation and BH merger, this metallicity evolution implies that the population of merging BHs traces the metallicity conditions of an earlier epoch rather than the merger redshift itself. As a consequence, the joint effect of metallicity evolution and time delays naturally induces a redshift dependence in the mass distribution of BBH. This evolution can be expressed through a window function $W_{t_{d}}(m;z_m)$ that maps the source-frame mass distribution at formation, $P_{s_1}(m;z_m)$, to the distribution observed at the merger redshift \citep{Mukherjee:2021rtw,Karathanasis:2022rtr}
\begin{equation}
    P(m;z_m) = P_{s_{1}}(m;z_m)\, W_{t_d}(m;z_m),
\end{equation}
where $z_m$ is the merging redshift, $W_{t_d}(m;z_m)$ can be expressed as

\begin{equation}
    W_{t_d}(m;z_m) = N\int_{z_m}^{\infty} P_{t_d}(t_d\mid t_d^{\text{min}}, d)\, \frac{dt}{dz'}\, W(m;z') \, dz',\label{wtd}
\end{equation}
where $ t_d^{\rm min}$ is the minimum time between formation of BHs and their merger, $N$ is the normalization constant, $W(m;z)$ is the Heaviside step function defined as $W(m;z_f) = \Theta(\mu_g(z_f)-m)$, and $z_f$ is the redshift of the formation of the BH.

We assume a linear dependence of the parameter $\mu_g$ of the mass distribution on the metallicity $Z$,
\begin{equation}
\mu_g(Z(z)) = \mu_g(Z_0) - \alpha_{\rm z}\,\log_{10}\left(Z(z)/{Z_0}\right),
\end{equation}
where $Z_0$ represents the median metallicity at $z=0$, and $\alpha_{\rm z}$ controls the strength of the metallicity dependence of $\mu_g(Z)$. The metallicity itself evolves with redshift, which we model as
\begin{equation}
\log Z(z) = \log Z_0 + \gamma_z\, z .
\end{equation}
Combining these expressions, we obtain
\begin{equation}
\mu_g(Z(z)) = \mu_g(Z_0) - \alpha_{\rm z}\, \gamma_{\rm z}\, z,
\label{eq:mu_z}
\end{equation}
where $\gamma_z$ characterizes the redshift evolution of the metallicity. Note that this model was tested before with the LVK GWTC-3 to provide the first measurement of the PISN \citep{farmer2019mind} mass-scale as well as the tentative signature of redshift-dependent mass distribution \cite{Karathanasis:2022rtr}.  {There are also studies which attempted to infer the redshift evolution of the BBH population from the GW catalog using different parametric and non-parametric models \cite{Rinaldi:2023bbd, Afroz:2025xpp, Gennari:2025nho}.}

\begin{figure}
  \subfigure[]{\label{fig:Rz}
    \centering    \includegraphics[width=\linewidth,trim={0.cm 0cm  0cm 0.cm},clip]{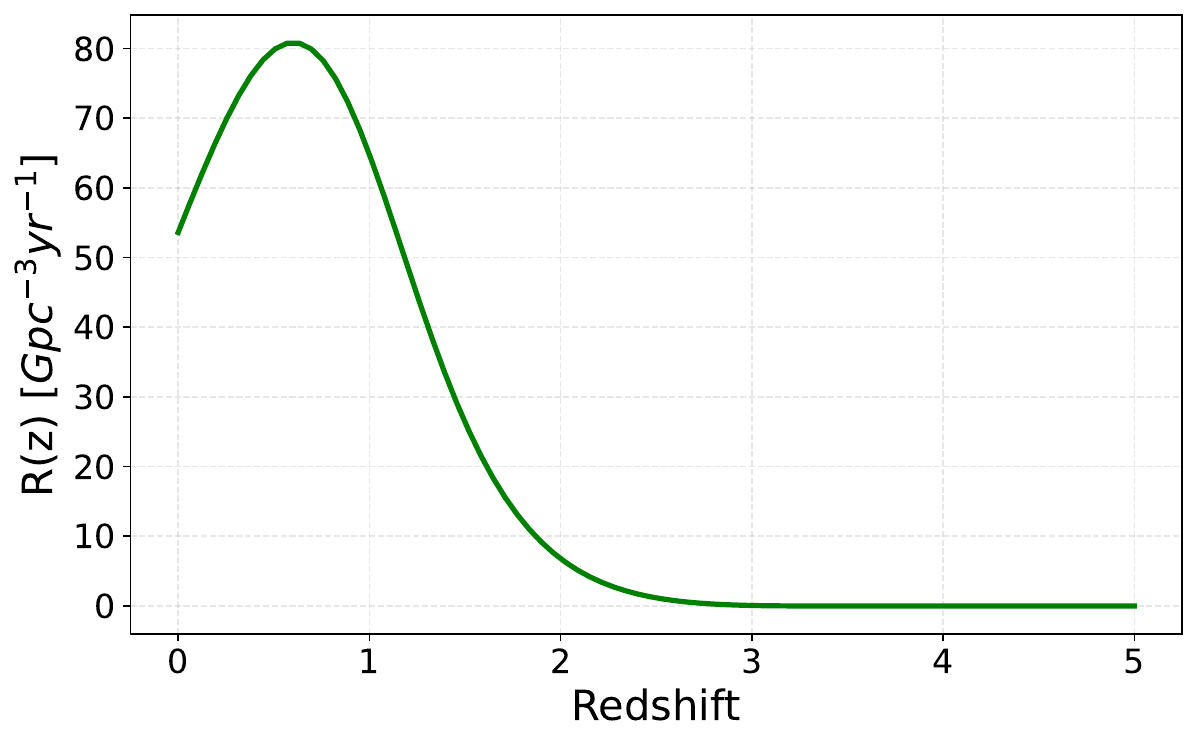}}
  \subfigure[]{\label{fig:Pz1}
    \centering    \includegraphics[width=\linewidth,trim={0.cm 0cm  0cm 0.cm},clip]{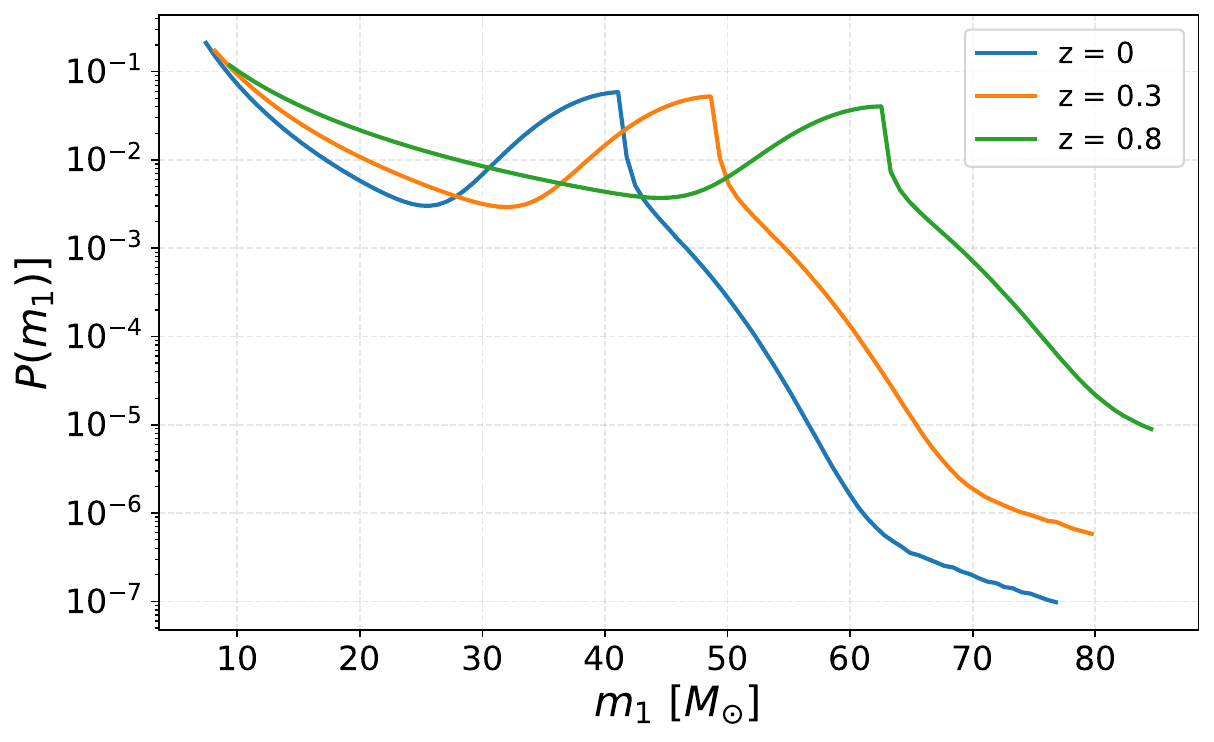}}
   \subfigure[]{\label{fig:Pz2}
    \centering
    \includegraphics[width=\linewidth,trim={0.cm 0  0 0.cm},clip]{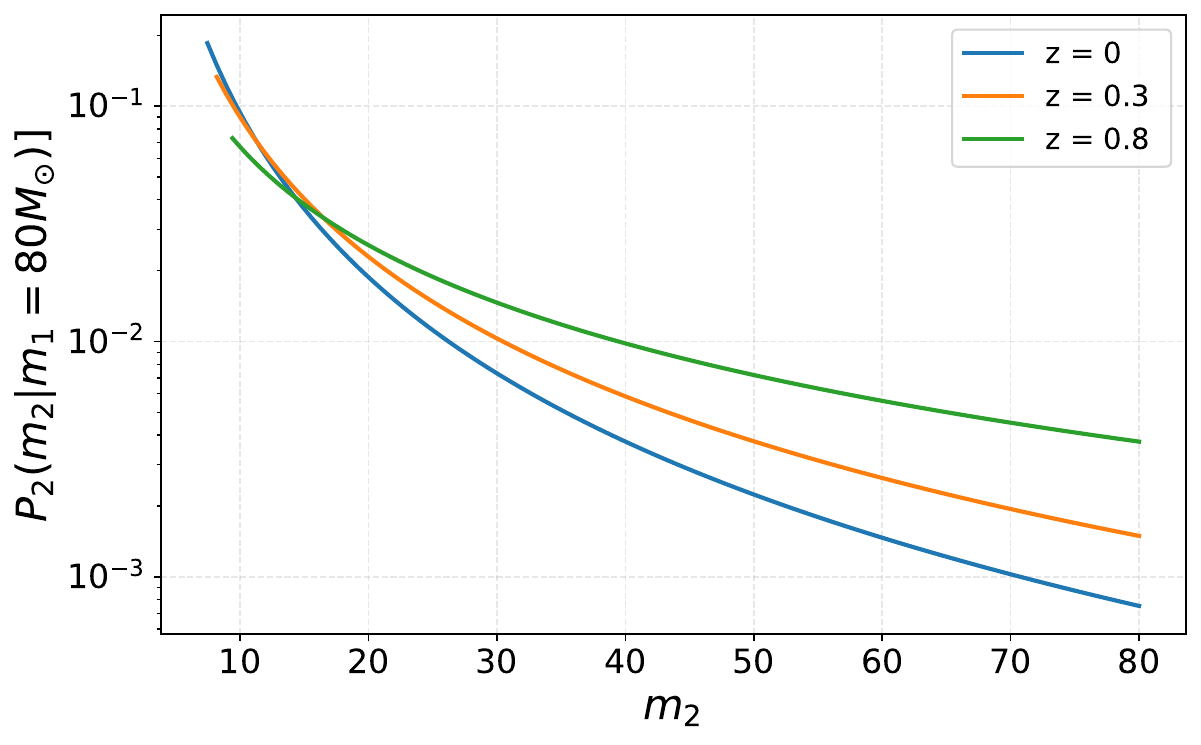}}
    \caption{(a) Merger rate as a function of redshift, (b) Primary mass distribution, and (c) Secondary mass distribution of the injected population with the hyperparameter values mentioned in Table \ref{tab:injections}. The merger rate is modeled as a function similar to the one used for the Madau-Dickinson star formation rate with a delay time distribution. The secondary mass distribution is described by a power-law function, whereas the primary mass distribution consists of a power-law term plus a Gaussian component whose peak evolves linearly with redshift.}
    \label{fig:Pop_model}
\end{figure}

In Fig. \ref{fig:Rz}, we show the merger-rate evolution as a function of redshift for the injected parameters. The model follows a Madau–Dickinson–like prescription with a time-delay distribution convolution, using the fiducial parameters summarized in Table \ref{tab:injections}. Due to the non-zero time delay, the merger rate shifts toward a lower redshift as compared to the formation rate of its progenitor. Similarly, Figs. \ref{fig:Pz1} and \ref{fig:Pz2} illustrate the primary mass distribution ($P(m_1)$) and secondary distribution ($P(m_2|m_1)$), respectively, of the merging BHs evaluated at several representative redshifts. These distributions reflect the mass models adopted for the injections: the secondary mass distribution is modeled as a power law, while the primary mass distribution is described by a combination of a power-law component and a Gaussian peak. As the redshift increases, the mass distributions shift toward higher masses, consistent with the injected redshift evolution of the parameters listed in Table \ref{tab:injections}. The curves at different redshifts illustrate a progressive increase in both the characteristic mass scale and the overall width of the distribution. Summarizing, the metallicity has two effects: one on $\mu_g$, and another on the mass distribution, through the function $W(m,z_f)$ in Eq.(\ref{wtd}).
On the other hand, the time delay affects the merger rate $R(z)$ and the single mass distribution, through Eq.(\ref{wtd}). The effects of the time delay and the metallicity are hence interconnected.

\begin{figure}
  \subfigure[]{\label{fig:Hist_dl}
    \centering    \includegraphics[width=\linewidth,trim={0.cm 0cm  0cm 0.cm},clip]{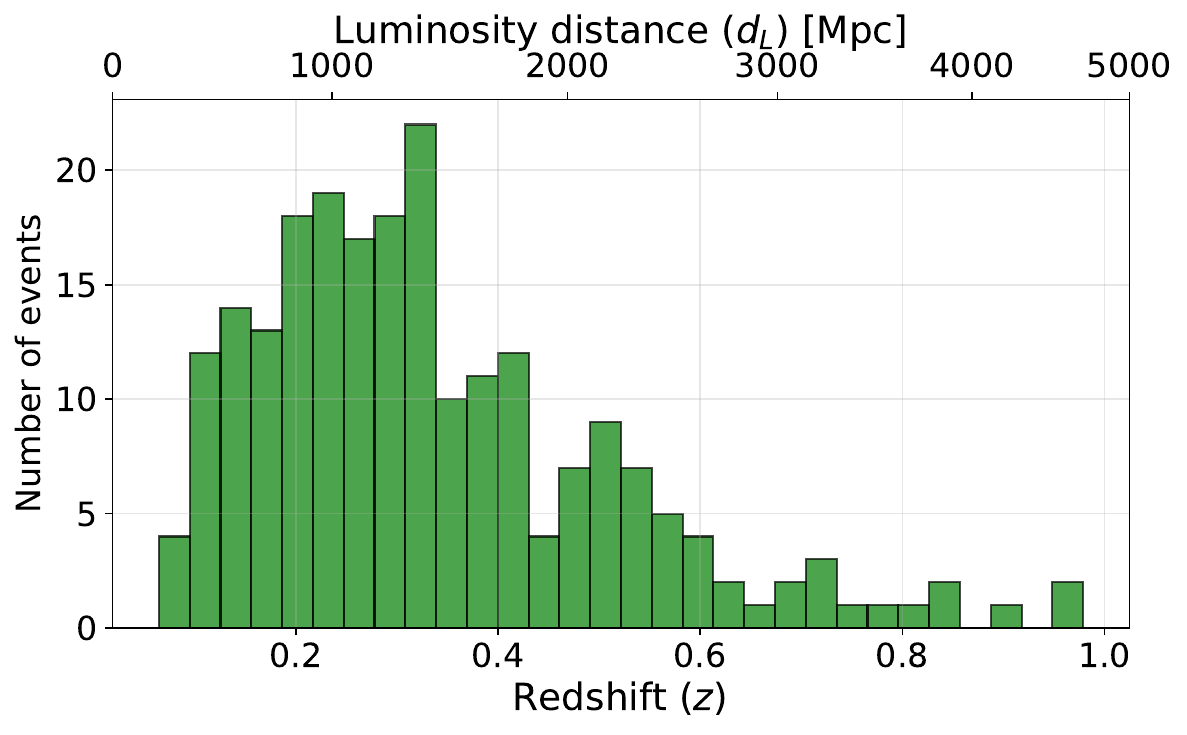}}
  \subfigure[]{\label{fig:Hist_m1}
    \centering    \includegraphics[width=\linewidth,trim={0.cm 0cm  0cm 0.cm},clip]{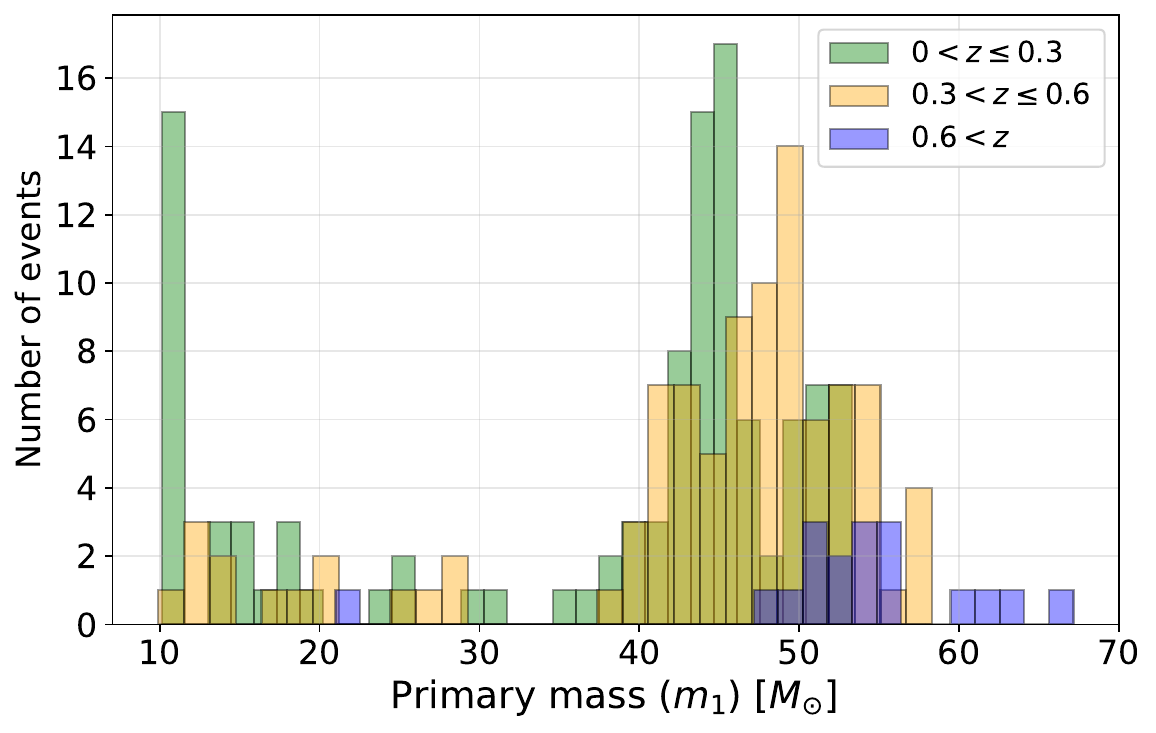}}
   \subfigure[]{\label{fig:Hist_m2}
    \centering
    \includegraphics[width=\linewidth,trim={0.cm 0  0 0.cm},clip]{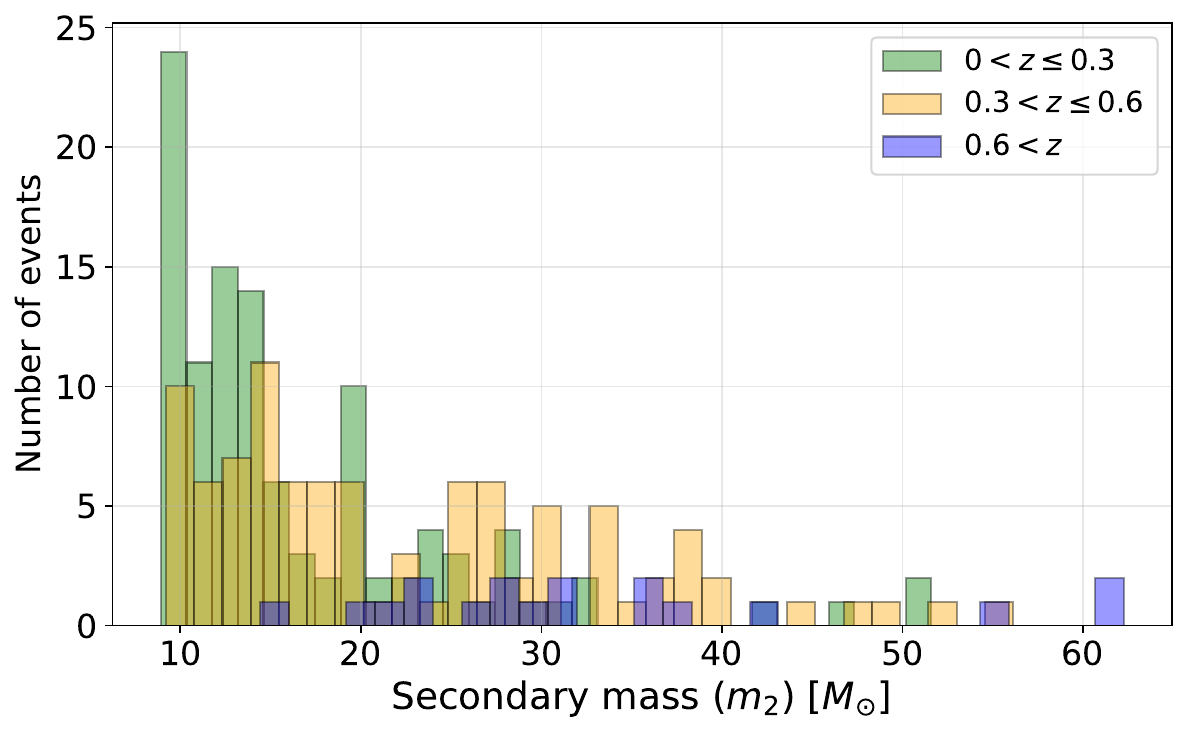}}
  \caption{Histogram of the number of detected events with an SNR threshold of $\rho_{\mathrm{th}} = 12$, assuming the O4 noise PSD. Panel (a) shows the redshift and luminosity distance distributions for the injected cosmology, while panels (b) and (c) display the source-frame primary and secondary BH mass distributions in three redshift bins, illustrating the redshift evolution of the mass scale.}
  \label{fig:Hist_m}
\end{figure}

Fig. \ref{fig:Hist_dl} shows the distribution of redshift and the corresponding luminosity distance of the detected events for the injected cosmological model. All the detected sources are below a luminosity distance of $ 5 ~ \mathrm{Gpc}$, reflecting the O4 detector sensitivity and the adopted SNR threshold. For an observing period of two years, this selection yields a total of 222 detected BBH events in our mock data set.  Figs. \ref{fig:Hist_m1} and \ref{fig:Hist_m2} present the histogram of the source-frame primary and secondary masses, respectively, for the detected events in three redshift bins. As expected, the BH masses increase with redshift, consistent with the positive redshift evolution of the characteristic mass scale described by Eq.~\eqref{eq:mu_z} for the injected model parameters. {It is important to note that the population model considered is still under simplistic assumptions. But in reality, the BBH population model can be complex due to various effects, to name a few: (i) variation of the initial stellar mass function with redshift and metallicity \cite{2023Natur.613..460L, Bate:2025oow}, (ii) metallicity-dependent SFR \cite{2010MNRAS.408.2115M, 2020A&A...636A..10C}, (iii) different astrophysical formation channels of BBHs \cite{Belczynski:2014iua,Mapelli:2021taw}, (iv) dependence of delay time distribution on metallicity and formation channel of BBHs \cite{Mandel:2015qlu,Boesky:2024wks}. Astrophysical variation of these quantities over cosmic time is large. Moreover, it becomes challenging to accurately model black hole formation from stars and predict mass distribution due to ignorance of stellar physics, such as the metallicity dependence of stellar winds \cite{2005A&A...442..587V, 2007A&A...473..603M}, core mass mixing due to stellar rotation \cite{2021A&A...655A..29J}, and nuclear reaction rates \cite{farmer2019mind}. As a result, the approximately $50\%$ variation with redshift is conservative, and in reality, this can be more complex.}

\begin{table*}
\caption{Summary of injected values of the population and cosmological parameters.}
\label{tab:injections}
\centering

\setlength{\tabcolsep}{10pt} 
\renewcommand{\arraystretch}{1.2} 

\begin{tabular}{l l r}
\hline\hline
\bf Parameter & \qquad \qquad \bf Description & \bf Injected Values \\
\hline\hline

\multicolumn{3}{c}{\texttt{$\Lambda$CDM Cosmological model}} \\
\hline
$H_0$ & Hubble constant in [km s$^{-1}$ Mpc$^{-1}$] & $92.87$ \\
$\Omega_{\rm m,0}$ & Matter energy density today & $0.3$ \\
\hline

\multicolumn{3}{c}{\texttt{Madau--Dickinson rate model}} \\
\hline
$R_0$ & Local merger rate in $\mathrm{Gpc^{-3}\,yr^{-1}}$ & $53.6$ \\
$\gamma$ & Power-law exponent of rate ($z \lesssim z_{\rm p}$) & $2.7$ \\
$\kappa$ & (Negative of) PL exponent of rate ($z \gtrsim z_{\rm p}$) & $2.9$ \\
$z_{\rm p}$ & Rate turnover redshift & $1.9$ \\
\hline

\multicolumn{3}{c}{\texttt{Time-delay parameters}} \\
\hline
$d$ & Spectral index of the delay-time distribution & $-1.28$ \\
$t_d^{\rm min}$ & Minimum delay time in Gyr & $1.841$ \\
$\alpha_{\rm z}\gamma_{\rm z}$ & Metallicity--redshift dependence of $\mu$ & $-19.6$ \\
\hline

\multicolumn{3}{c}{\texttt{Power-law + Gaussian peak mass model}} \\
\hline
$\alpha$ & PL index of primary mass & $3.64 - 1.53\,z$ \\
$\beta$ & PL index of secondary mass & $-2.32 + 1.167\,z$ \\
$m_{\rm min}$ & Minimum source mass [$M_\odot$] & $7.45 + 2.44\,z$ \\
$m_{\rm max}$ & Maximum source mass [$M_\odot$] & $76.79 + 9.72\,z$ \\
$\delta_m$ & Low-mass smoothing scale [$M_\odot$] & $4.80$ \\
$\mu_g$ & Peak of the Gaussian [$M_\odot$] & $41.58 - \alpha_{\rm z}\gamma_{\rm z} z$ \\
$\sigma_g$ & Gaussian width [$M_\odot$] & $5.33 + 1.859\,z$ \\
$\lambda_{\rm peak}$ & Fraction in Gaussian component & $0.393 - 0.021\,z$ \\
\hline\hline
\end{tabular}
\end{table*}

\newpage

\section{Methods}\label{sec:methods}

As introduced in Sec. \ref{sec:setup}, the code \icarogw{} ~\citep{Mastrogiovanni:2023zbw} is used to jointly constrain cosmological and population parameters $\Lambda$ from the simulated BBHs events data (all the parameters are briefly described in Tab. \ref{tab:injections}). We use the same method as in the first MDC paper, namely the spectral siren method. We also assume an astrophysical origin for BHs, and hence we use a function similar to the one used to fit the SFR \citep{Madau:2014bja} to model the BBHs merger rate $R(z)$. Contrary to the simulated BBHs event merger rate model, we do not take into account the delay time distribution between formation and merger time of the binaries for our analyses.  Therefore, we use the same Madau-Dickinson parametric form for the BBHs merger rate as in the analyses of the first MDC paper, which is captured by the parameters $\gamma$, $\kappa$, $z_p$, and $R_0$, where $R_0$ is the merger rate of BBHs at $z=0$.
Similarly, we assume the same redshift-independent source frame mass distributions as in the analyses of the first \texttt{Blinded-MDC} study. Thus, we adopt PLG and PL distributions for the primary mass $(m_1)$ and the secondary mass $(m_2)$, respectively. Finally, we generate posterior samples for the cosmological and population parameters $\Lambda$ using \bilby{}'s \citep{bilby_paper} sampler \texttt{dynesty} \citep{dynesty_paper} within the \icarogw{} package \cite{Mastrogiovanni:2023zbw}. 

\subsection{Spectral sirens method: \icarogw{} setup}

The joint posterior of the cosmological and population parameters $\Lambda$, given $N_{\rm obs}$ GW observations, each described by individual source parameters $\theta$, with data $\{x\}$ for a given observation time $T_{\rm obs}$, is given by (\cite{Mandel:2018mve, Vitale:2020aaz, Mastrogiovanni:2021wsd, Mastrogiovanni:2023zbw})
\begin{equation}
    p(\Lambda |\{x\})\propto \pi (\Lambda) {e^{ - {N_{\exp }}(\Lambda )}}\prod\limits_{i = 1}^{{N_{{\rm{obs }}}}} {{T_{{\rm{obs }}}}} \int {\cal L} \left( {{x_i}\mid \theta ,\Lambda } \right){{{\rm{d}}N} \over {{\rm{d}}t{\rm{d}}\theta }}(\Lambda ){\rm{d}}\theta ,
\label{eq:combined_posterior}
\end{equation}
where $\pi (\Lambda)$ is the prior on both cosmological and population parameters (see Tab. \ref{tab:priors}). The product across the $N_{\rm obs}$ observations is because each GW event is independent of the others. The two key quantities to evaluate in Eq. \ref{eq:combined_posterior} are the "expected number of detected events", $N_{\exp }(\Lambda)$, and the integral of the product between the likelihood of a single event, ${\cal L} \left( {{x_i}\mid \theta,\Lambda } \right)$, and the production rate of CBC events, ${{{\rm{d}}N} \over {{\rm{d}}t{\rm{d}}\theta }}(\Lambda )$.

The expected number of detected events accounts for the selection bias and is given by
\begin{equation}
    {N_{{\rm{exp }}}}(\Lambda ) = {T_{{\rm{obs }}}}\int_{\rho\geqslant \rho_\text{th}} {{p_{{\rm{det }}}}} (\theta ,\Lambda ){{{\rm{d}}N} \over {{\rm{d}}t{\rm{d}}\theta }}{\rm{d}}\theta ,
\label{eq:Nexp_integral}
\end{equation}
where $p_{\rm{det }} (\theta ,\Lambda )$ is the detection probability. However, in general, we are not provided with an analytical form of this detection probability. Therefore, the selection bias is computed in \icarogw{} using the current approach based on a large set of $N_{\rm{gen }}$ simulations of injected events, which are usually referred to as "injections", generated from a prior $\pi _{\rm{inj }}(\theta)$. The occurrence of $N_{\rm{det }}$ detected injections with a matched filtering SNR $\rho$ above a given fixed SNR threshold $\rho_{\rm{th}}$, which is proportional to $p_{\rm{det }} (\theta ,\Lambda )$, allows us to compute the integral in Eq. \ref{eq:Nexp_integral} via Monte Carlo integration according to \cite{agarwal2025blinded}
\begin{equation}
    {\left. {{N_{{\rm{exp }}}} \approx {{{T_{{\rm{obs }}}}} \over {{N_{{\rm{gen }}}}}}\sum\limits_{j = 1}^{{N_{{\rm{det }}}}} {{1 \over {{\pi _{{\rm{inj }}}}\left( {{\theta _j}} \right)}}} {{{\rm{d}}N} \over {{\rm{d}}t{\rm{d}}\theta }}} \right|_j}.
\label{eq:Nexp_MCIntegral}
\end{equation} 
This Monte Carlo integration is performed only for detected injections that pass the same SNR threshold criteria as the detected mock events ($\rho_{\mathrm{th}} = 12$ in this analysis). 
The effect of the number and realization of the injection samples on the estimation of population and cosmological parameters was studied in the first MDC paper\cite{agarwal2025blinded}. It was found that the evaluation of the Monte Carlo sum was partially impacted by the number of injections, but it was not affected by the specific realization when the sample is sufficiently large ($\sim 10^5$ injections) for the Monte Carlo sum to converge.


\subsection{Priors settings}

We consider broad and non-informative priors on 14 hyperparameters described in Tab. \ref{tab:priors}. Note that these priors correspond to the same priors used for the redshift-dependent scenario analyses of the first \texttt{blinded-MDC} study. 

\begin{table}
\caption{List of priors used in the analyses for the Redshift-Dependent scenario.}
\label{tab:priors}
\centering
\begin{tabular}{l r}
\multicolumn{1}{c}{\bf Parameter} & \multicolumn{1}{c}{\bf Priors}\\
\hline
\hline
$H_0$ [km s$^{-1}$ Mpc$^{-1}$] &$\mathcal{U}(10, 250)$\\
$\Omega_{\rm m, 0}$ &$\mathcal{U}(0.1, 0.9)$\\
\hline
\hline
$\alpha$ &$\mathcal{U}(1.5, 12)$\\
$\beta$ &$\mathcal{U}(-4, 12)$\\
$m_{\rm min}$ [$M_{\odot}$] &$\mathcal{U}(2, 10)$\\
$m_{\rm max}$ [$M_{\odot}$] &$\mathcal{U}(50, 200)$\\
$\delta_{\rm m}$ [$M_{\odot}$] &$\mathcal{U}(0, 10)$\\
$\mu_{\rm g}$ [$M_{\odot}$] &$\mathcal{U}(10, 80)$\\
$\sigma_{\rm g}$  [$M_{\odot}$] &$\mathcal{U}(0.4, 20)$\\
$\lambda_{\rm peak}$ &$\mathcal{U}(0, 1)$\\
\hline
\hline
$\gamma$ &$\mathcal{U}(0, 12)$\\
$\kappa$  &$\mathcal{U}(0, 6)$\\
$z_{\rm p}$ &$\mathcal{U}(0, 4)$\\
$R_0$ [Gpc$^{-3}$ yr$^{-1}$] &$\mathcal{\log U}(10^{-2}, 10^3)$\\
\hline
\end{tabular}
\end{table}

\section{Results from the blinded-MDC}\label{sec:results}

We use \icarogw{} to jointly estimate cosmological and population parameters from 222 BBH events detected during two years of observing time. In order to understand the impact of the observing time and the number of detected events on parameter inference, we also perform separate analyses on the 109 and 113 detected events during the first and second years of observation, respectively. All of our analyses are performed using the methods and modeling assumptions described in Sec. \ref{sec:methods}. The probability of detection is computed using a set of $10^5$ injections for all the analyses. 

\begin{figure}
    \includegraphics[scale = 0.45]{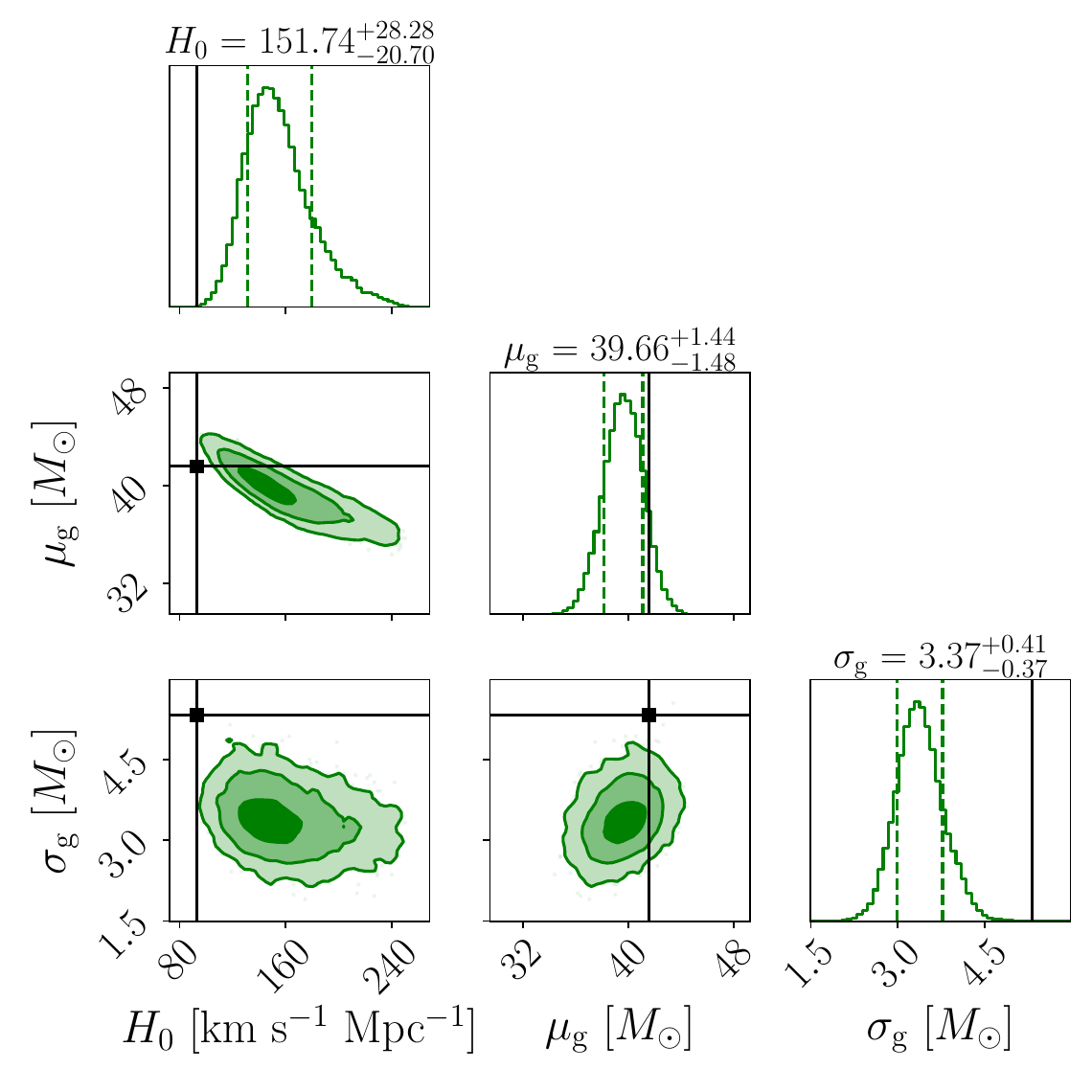}
    \caption{Reduced corner plot of $H_0$, $\mu_g$, and $\sigma_g$ inferred from the analysis of 222 detected events during 2 years of observing time. The probability of detection is computed using a set of $10^5$ injections. The horizontal and vertical solid lines indicate the injected (true) values of the parameters at $z=0$. The shaded contours represent the $39.3\%$, $86.5\%$, and $98.9\%$ credible regions of the marginalized 2D posteriors. The vertical dashed lines denote the $68\%$ credible interval of the marginalized 1D posteriors. The corner plot for all parameters is shown in Fig.(\ref{fig:full_corner_2years}). }
    \label{fig:red_corner_2years}
\end{figure}

Fig. \ref{fig:red_corner_2years} shows a reduced corner plot with the two-dimensional and one-dimensional projections of the posterior samples of the Hubble constant, $H_0$, and two population parameters, $\mu_g$ and $\sigma_g$, that strongly correlate with $H_0$. The shaded regions in the two-dimensional projections represent the $39.3\%$, $86.5\%$, and $98.9\%$ credible regions, which correspond to the boundaries at distances of $1\sigma$, $2\sigma$, and $3\sigma$ from the mean of the marginalized posterior if the marginalized posteriors were perfect 2D Gaussian distributions. The vertical dashed lines in the one-dimensional histograms on the diagonal encompass the $68\%$ credible interval. The horizontal and vertical solid lines indicate the injected values of the parameters, which, for the redshift-dependent parameters $\mu_g$ and $\sigma_g$, correspond to their values at $z=0$. From this figure, we can clearly see a systematic bias due to the mismatch between the astrophysical population model used for the analysis and the true population model that was used to simulate the BBH mock events. Crucially, the true injected value of $H_0$ is located near the edge of the posterior range of this parameter. The complete corner plot showing all the distributions of cosmological and population parameters is presented in Fig. \ref{fig:full_corner_2years} of the Appendix \ref{app:full_posteriors}. While $\Omega_{m,0}$ is also inferred, as can be seen in this plot, our analysis exhibits limited sensitivity to this parameter, with the 1D posterior predominantly shaped by the prior range.

\begin{figure}
    \includegraphics[scale = 0.5]{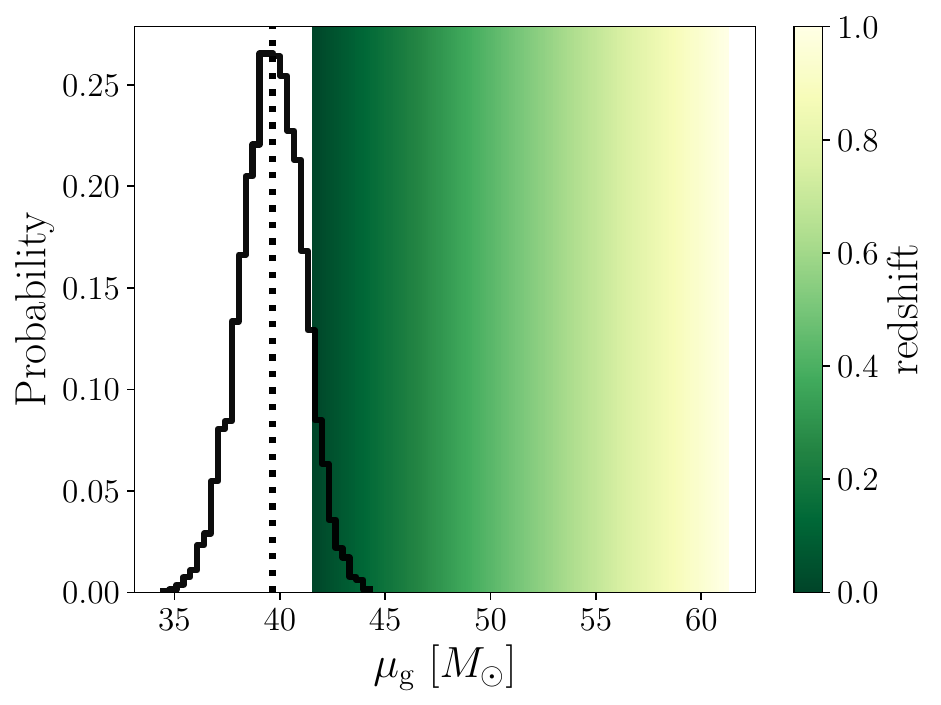}
    \caption{Posterior samples obtained for $\mu_g$ are plotted in black. The vertical dotted line corresponds to the median of the distribution. The colored plot shows the injected values as a function of redshift according to $\mu_g(z) = 41.58 + 19.65 z$.}    \label{fig:mu_g_posterior}
\end{figure}

\begin{figure}
    \includegraphics[scale = 0.5]{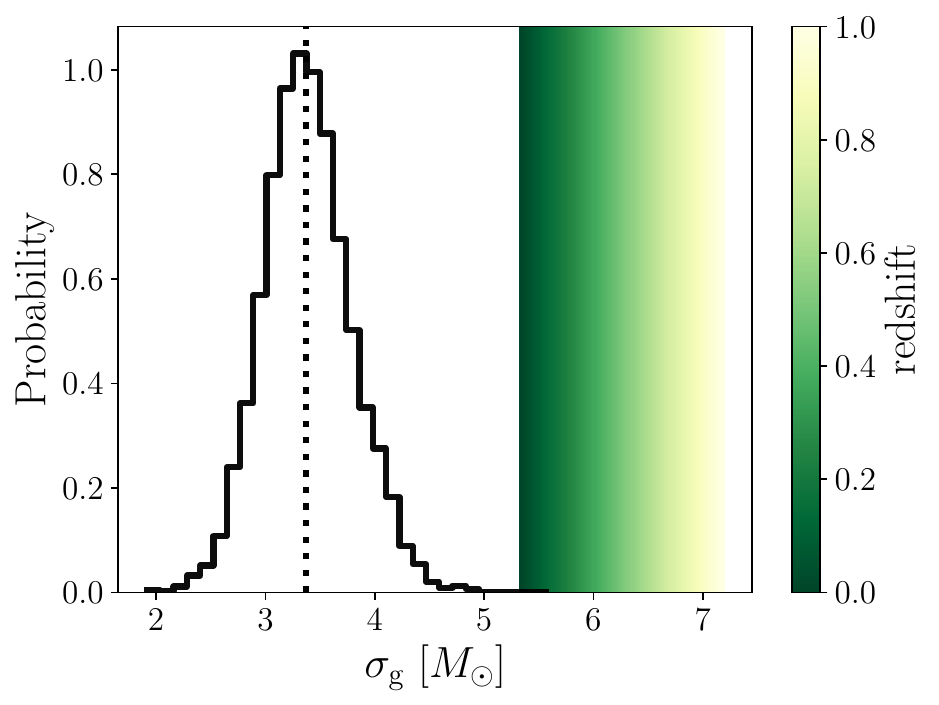}
    \caption{Posterior samples obtained for $\sigma_g$ are plotted in black. The vertical dotted line corresponds to the median of the distribution. The colored plot shows the injected values as a function of redshift according to $\sigma_g(z) = 5.33 + 1.859 z$.}    \label{fig:sigma_g_posterior}
\end{figure}

\begin{figure}
    \includegraphics[scale = 0.5]{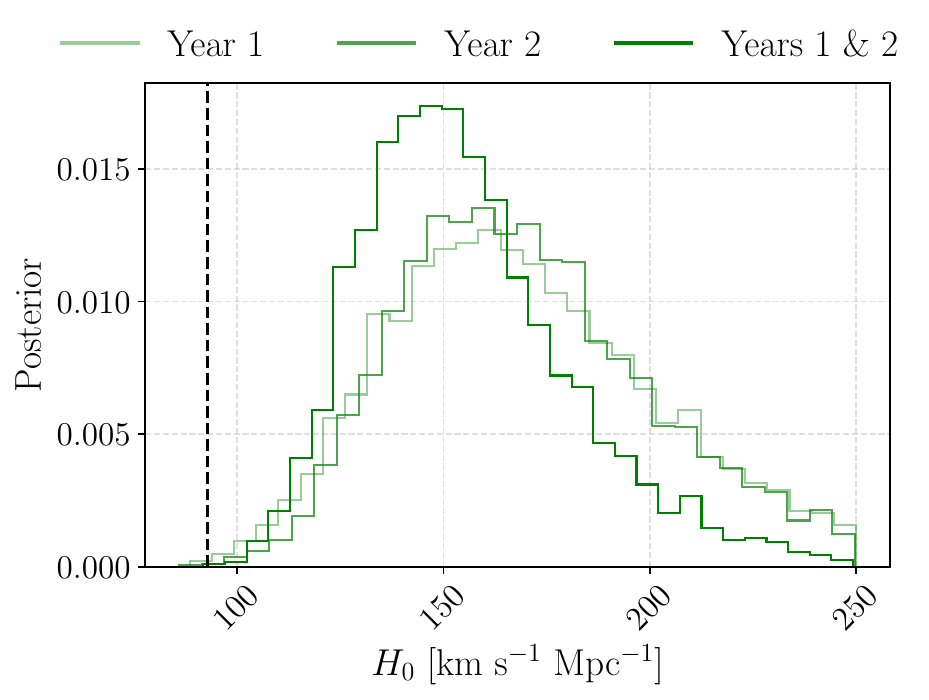}
    \caption{Impact of observing time and number of detected events on $H_0$ estimation. Marginalized 1D posterior samples obtained from three distinct analyses: Year 1 only events (109 events), Year 2 only events (113 events), and the cumulative events from both years (222 events). The vertical dashed line corresponds to the injected (true) value.}    \label{fig:H0_posterior_ev_per_year}
\end{figure}

\begin{figure*}
\centering
    \includegraphics[scale = 0.5]{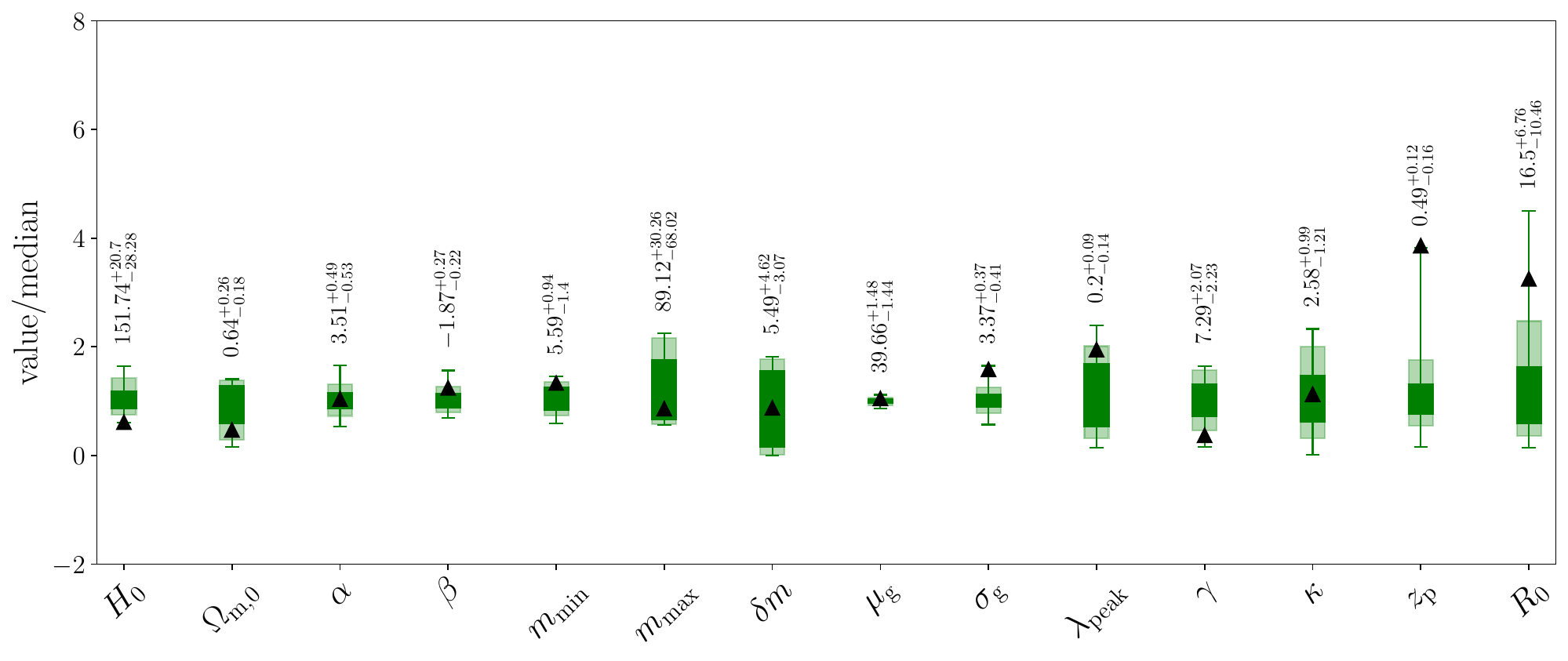}
    \caption{Summary plot of the results obtained for the scenario investigated in this paper. We used all the detected events, 222 during two years of observing time, and $10^5$ injections to compute the selection effects. The horizontal axis corresponds to the parameters estimated, and the vertical axis displays the normalized marginalized 1D posterior distributions. The posterior samples have been divided by the median of their respective distribution, which allows a direct visual comparison of the relative uncertainty across all parameters in a single figure. The dark and light shaded boxes encompass the $68\%$ and $95.4\%$ credible intervals around the median, and the whiskers stand for the entire distribution's range. The black triangles correspond to the injected (true) values, divided by the corresponding median of each distribution. The texts next to the boxplots of the parameters give their median and $68\%$ credible interval.  }    \label{fig:summary_plot}
\end{figure*}

In Figs. \ref{fig:mu_g_posterior} and \ref{fig:sigma_g_posterior}, we compare the inferred marginal distributions of $\mu_g$ and $\sigma_g$, respectively, with their true injected values at different redshifts  (shown by the color-bar). These plots show that the injected values of the parameters at different redshifts fall outside the $68\%$ credible interval of their posterior distributions. In contrast, the results of the previous \texttt{Blinded-MDC} study demonstrated that the injected values of these two $z$-dependent parameters of the true mass distribution were accurately recovered (i.e., falling within the $68\%$ credible interval of their posterior distributions) when the merger rate was modeled without a time delay distribution for both the simulation and the analysis of mock events. Crucially, while both the first MDC study analysis and the one presented in this paper mis-model the mass distribution by assuming a redshift-independent model, the discrepancy observed in this new scenario (the biased recovery of $\mu_g$ and $\sigma_g$) arises from two effects, (i) a non-zero value of delay-time distribution leads to mixing of black holes from different formation redshifts which spans over a range of metallicity, (ii) variation of metallicity leads to a variation in the PISN mass scale. The interplay between these two effects produces a BBH population, which shows a higher $\mu_g$ for more delayed mergers. As a result, an analysis with a model that does not capture this aspect ends up inferring a different value of the population parameters due to mis-modelling. This mis-modelling leads to a shift in the peak of the posterior distribution of the Hubble constant, from the injected value in the simulation as shown in Fig. \ref{fig:H0_posterior_ev_per_year}. As we can see from this plot, the analyses of the first and second years of observations yield similar results, while the analysis of the entire set of events detected during the two years of observations provides more robust constraints. We also summarize the comparison of the injected values of the parameters with their marginalized posterior distributions in Fig. \ref{fig:summary_plot}. The shaded boxes represent the 1-$\sigma$ and 2-$\sigma$ credible intervals, and the whiskers stand for the entire distribution's range. 

\subsection{Summary on the impact of black hole population assumption on the Hubble constant inference}
The \texttt{Blinded-MDC} performed in this analysis sheds light on a key understanding of how mis-modelling of the astrophysical population of the BBHs can lead to a wrong inference of the cosmological parameters. As one of the primary science goals of the GW cosmology is the inference of the value of the expansion history of the Universe, we list down the key findings as well as the requirements for the spectral siren method to work for inferring a reliable value of the expansion history.

\begin{itemize}
    \item The spectral sirens method, which uses the BBH mass distribution to infer the value of the expansion history of the Universe, gets impacted by the mis-modelling of the astrophysical BBH mass distribution, and untested assumptions on the redshift evolution of the BBH mass distribution.
    \item The impact of stellar metallicity evolution with redshift drives the BBH mass distribution. As the stellar metallicity decreases with redshift, the  BBH mass distribution may not remain independent of redshift, and  {hence the spectral siren technique can be impacted due to degeneracy between population models and cosmological inferences arising due to the spectral shape of the mass distribution.} 
    \item This \texttt{Blinded-MDC} finds that such scenarios of redshift evolution of BBH mass distribution, even only within about $40-50\%$ in the redshift range $z \in [0,1]$ for the parameters (as mentioned in the table \ref{tab:injections}) which controls the mass distribution, will cause a bias in the value of the inferred value of the Hubble constant by about $60\%$. 
    \item In order to obtain an accurate and precise independent measurement of the Hubble constant using GW sources, which can shed light on about $8\%$ discrepancy in the value of the Hubble constant (which is usually referred to as the Hubble tension \cite{Verde:2019ivm, DiValentino:2021izs}), the Hubble measurement needs to be made with a precision of about a few percent (say about $\sim 2\%$ statistical uncertainty\footnote{In order to reach a conclusive understanding of the Hubble tension indicating a difference in the value of the Hubble constant between CMB \cite{Planck:2018vyg} and SHOES measurement\cite{Riess:2019qba}, one needs to make about a $\sim 2\%$ measurement of Hubble constant which can shed light on the Hubble tension at close to $5\sigma$.}). This would need any systematic error on the mass spectrum to be much less than $2\%$ statistical uncertainty in order to have a reliable inference. 
    \item For methods using the spectral siren technique, as a result,  {in order to resolve the Hubble tension, which is at the level of $8\%$, the mass spectrum of the BBHs needs to be known at a percent level accuracy from an independent method (which is independent of cosmological model)}. Otherwise, the degeneracy between BBH population models can map the observed data into an incorrect parameter space of BBH distribution and hence can bias the redshift inference of the sources, leading to a biased value of the Hubble constant\footnote{The features in the BBH mass spectrum that strongly correlate with the Hubble constant can drive maximum bias in the inferred value.}. 
\end{itemize}

\section{Conclusion}\label{conc}
The \texttt{Blinded-MDC} performed in this analysis tests the robustness of the spectral siren method, which is used to infer the expansion history of the Universe by assuming non-evolution of the BBH mass spectrum. We show that such an untested assumption in the analysis makes the inference of the current expansion rate, the Hubble constant $H_0$, unreliable. We show that as the metallicity in the Universe decreases with increasing redshift, it can drive an evolution of the BBHs mass distribution with redshift. Moreover,  as there is an intrinsic uncertainty in the parent star metallicity, its stellar wind dependence on the remnant black hole mass, and multiple formation channels of BBHs, it becomes significantly challenging to make a robust prediction of the BBH mass distribution across cosmic redshifts. As a result, GW cosmology methods such as spectral sirens, which are based on untested assumptions of non-evolving mass distribution, can be impacted. The previous \texttt{Blinded-MDC}, which only allowed a linear increase in the mass distribution and no dependency on the delay time, also led to incorrect inference towards a lower value of the Hubble constant than the injected one \cite{agarwal2025blinded}. This new MDC sheds light on the interplay between the Hubble constant, with both time delay and mass distribution, which can drive an inaccurate inference when the underlying astrophysical population model is mis-modelled.  {However, it is important to note that the particular amount of impact will depend on the true underlying astrophysical model and also on the degree of redshift evolution. If the degree of redshift evolution remains negligible in comparison to the statistical uncertainty of the measurement, then the impact from population mis-modeling will be less significant.}  

We show that in a setup with O4 detector noise for three detector configurations, LIGO-Hanford, LIGO-Livingston, and Virgo that for a model with redshift evolving mass distribution of the BBHs towards a higher value by about $40\%$ in the redshift range $z \in [0-1]$, there is bias in the inference of the Hubble constant $H_0$ by about $60\%$ towards a higher side. This bias is attributed to incorrect inference of the BBH source redshifts arising from the mis-modelling of the mass distribution and merger rate. As the underlying astrophysical model considered in the \texttt{Blinded-MDC} differs from the assumed astrophysical model in the analysis, the resultant values of the astrophysical population are not recovered correctly, and also the inferred redshift of the GW sources becomes incorrect. Hence, an incorrect value of redshift inference leads to an incorrect inference of the Hubble constant. Though the testing of the robustness of the spectral siren technique performed in this \texttt{Blinded-MDC} is for a particular astrophysical model involving both metallicity and delay time distribution dependent BBH mass distribution and merger rate, the key summary remains the same:  {that an untested assumption on the astrophysical BBH population is vulnerable to induce systematic errors in the Hubble constant inference.} 

The fact that the standard-siren technique aims to provide one of the robust inferences of the Hubble constant, with the key advantage that such measurements do not need an external calibration, suggests that the reliability of the technique should be tested. This \texttt{Blinded-MDC} tests the susceptibility of the spectral-siren technique to the ignorance of the mass distribution. Though different assumptions can be made for the parametric and non-parametric reconstruction of the mass distribution, the degeneracy in the astrophysical model parameters and cosmological model parameters can obscure the inference of the true Hubble constant in a robust way unless the astrophysical population of the BBHs is independently measured accurately. In order to have an accurate inference of the Hubble constant, which can resolve the Hubble tension, the systematic uncertainty in the Hubble constant needs to be less than the statistical error. So to obtain a $2\%$ measurement of the Hubble constant from GW sources, which can shed light on the $\sim 8\%$ discrepancy in the value of $H_0$ between SH0ES and CMB, the systematic uncertainty on the BBH mass distribution should be smaller than $2\%$. It is important to note, as found from this study, even though the BBH population parameters are consistent with each other at different redshift bins and do not show any statistically significant deviation (due to large uncertainties), it can still be at a wrong value, due to an incorrect underlying model assumption. This is the key reason for the incorrect value of $H_0$. In the future, it will be pertinent to understand the budget of systematic error on $H_0$ due to the BBH mass distribution from GW data for making a reliable inference of the expansion history of the Universe.  
\section*{Acknowledgements}
The authors are very thankful to Gergely Dalya for reviewing the manuscript as part of the LIGO Publication and Presentation policy and providing valuable comments on the draft. The authors thank the LIGO-Virgo-KAGRA Scientific Collaboration for providing noise curves. The authors are grateful for computational resources provided by the LIGO Laboratory and supported by National Science Foundation Grants PHY-0757058 and PHY-0823459 and the computing resources of the \texttt{⟨data|theory⟩ Universe-Lab}, supported by TIFR and the Department of Atomic Energy, Government of India. LIGO, funded by the U.S. National Science Foundation (NSF), and Virgo, supported by the French CNRS, Italian INFN, and Dutch Nikhef, along with contributions from Polish and Hungarian institutes. This collaborative effort is backed by the NSF’s LIGO Laboratory, a major facility fully funded by the National Science Foundation. The research leverages data and software from the Gravitational Wave Open Science Center, a service provided by LIGO Laboratory, the LIGO Scientific Collaboration, Virgo Collaboration, and KAGRA. Advanced LIGO's construction and operation receive support from STFC of the UK, Max-Planck Society (MPS), and the State of Niedersachsen/Germany, with additional backing from the Australian Research Council. Virgo, affiliated with the European Gravitational Observatory (EGO), secures funding through contributions from various European institutions. Meanwhile, KAGRA's construction and operation are funded by MEXT, JSPS, NRF, MSIT, AS, and MoST. This material is based upon work supported by NSF’s LIGO Laboratory which is a major facility fully funded by the National Science Foundation. This work is supported in part by the Perimeter Institute for Theoretical Physics. Research at Perimeter Institute is supported by the Government of Canada through the Department of Innovation, Science and Economic Development Canada and by the Province of Ontario through the Ministry of Economic Development, Job Creation and Trade. The work of SM and MRS is part of the \texttt{⟨data|theory⟩ Universe-Lab}, supported by TIFR and the Department of Atomic Energy, Government of India.  AER was supported by the UDEA projects
 2021-44670, 2019-28270, 2023-63330. JGB acknowledges support of the Spanish Agencia Estatal de Investigaci\'on through the Research Project PID2024-159420NB-C43 [MICINN-FEDER], and the grant ``IFT Centro de Excelencia Severo Ochoa CEX2020-001007-S. 
 
\bibliographystyle{unsrt}
\bibliography{main_arxiv.bib}{}

\appendix

\section{Full cosmological and population parameters posteriors}
\label{app:full_posteriors}

\begin{figure*}
    \centering
    \includegraphics[scale = 0.24]{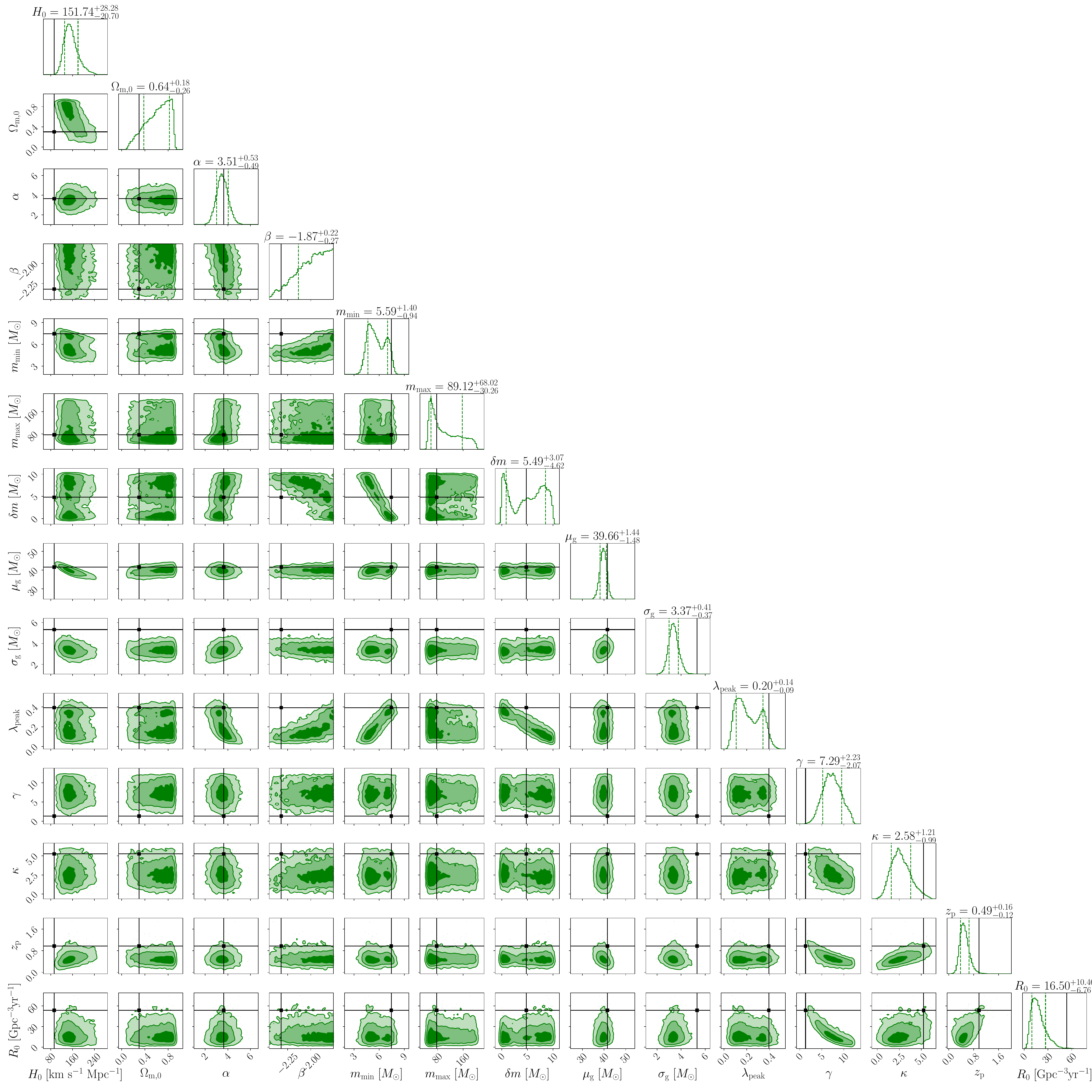}
    \caption{Posterior samples inferred from the analysis of 222 detected events during 2 years of observing time. The probability of detection is computed using a set of $10^5$ injections. The horizontal and vertical solid lines indicate the injected (true) values of the parameters at $z=0$. The shaded contours represent the $39.3\%$, $86.5\%$, and $98.9\%$ credible regions of the marginalized 2D posteriors. The vertical dashed lines denote the $68\%$ credible interval of the marginalized 1D posteriors.}
    \label{fig:full_corner_2years}
\end{figure*}

In this appendix, we report the corner plot for all cosmological and population parameters, which is shown in Fig.(\ref{fig:full_corner_2years}). We used all the detected events, 222 during two years of observing time, and $10^5$ injections to compute the selection effects.  

\end{document}